\documentclass[12pt,preprint]{aastex}
\usepackage{epsfig}
\usepackage{rotating}
\usepackage{graphicx}
\def\xmm{{\sl XMM}-Newton}
\def\fd{\dot{f}}

\begin{document}

\shortauthors{Zavlin, Pavlov, \& Sanwal}
\shorttitle{Timing properties of 1E~1207.4--5209}
\title{
Variations in the spin period of the radio-quiet pulsar 1E~1207.4--5209
}
\author{Vyacheslav E.\ Zavlin\altaffilmark{1}\altaffiltext{1}{
Max-Planck Institut f\"ur Extraterrestrische Physik, 85748 Garching, 
Germany;
zavlin@mpe.mpg.de},
George G.\ Pavlov\altaffilmark{2}\altaffiltext{2}{
Dept.\ of Astronomy and Astrophysics, The Pennsylvania State
University, 525 Davey Lab, University Park, PA 16802; 
pavlov@astro.psu.edu, divas@astro.psu.edu}, and
Divas Sanwal\altaffilmark{2}
}
\begin{abstract}
The X-ray source 1E~1207.4$-$5209 is a compact central object
in the G296.5+10.0 supernova remnant. Its spin period of 424 ms, discovered
with the {\sl Chandra} X-ray Observatory, suggests that it is a neutron star. 
The X-ray spectrum of this radio-quiet pulsar shows at least 
two absorption lines,
first spectral features discovered in radiation from an isolated neutron star.
Here we report the results of timing analysis of {\sl Chandra} and \xmm\
observations of this source showing a non-monotonous behavior of its period. 
We discuss three hypotheses which may explain the observational
result.  The first one assumes that 1E~1207.4$-$5209 is a glitching pulsar, 
with frequency jumps of
$\Delta f\gtrsim 5\, \mu$Hz occurring every 1--2 years. 
The second hypothesis explains the deviations from a steady spin-down
as due to accretion, with accretion rate varying from $\sim 10^{13}$
to $\gtrsim 10^{16}$ g s$^{-1}$, from a disk
possibly formed from ejecta produced in the supernova explosion.
Finally, the period variations could be explained assuming that the pulsar
is in a wide binary system with a long period, 
$P_{\rm orb} \sim 0.2$--6 yr, and a low-mass companion, $M_2< 0.3 M_\odot$.
\end{abstract}
\section{Introduction}
X-ray observations of supernova remnants (SNRs) have revealed a number of
radio-quiet central compact objects (CCOs --- see Pavlov et al.\ 
2002a, 2004
for a review) whose nature remains enigmatic. They are characterized by soft,
apparently thermal, X-ray spectra and a lack of manifestations of pulsar
activity (e.g., radio and/or $\gamma$-ray emission, 
compact synchrotron nebulae).
Most likely, CCOs are neutron stars (NSs) formed in supernova explosions.

One of the best investigated CCOs is 1E 1207.4--5209 (1E1207 hereafter) in the 
G296.5+10.0 (= PKS 1209--51/52) SNR.  It was discovered by Helfand \& Becker
(1984) with the {\sl Einstein} observatory. Mereghetti, Bignami \& Caraveo
(1996) and Vasisht et al.\ (1997) interpreted the {\sl ROSAT} and {\sl ASCA}
spectra of 1E1207 as blackbody emission of 
a (redshifted) temperature $kT^\infty\simeq 0.25$~keV from an area
with radius $R^\infty\simeq 1.5$~km at $d=2$ kpc.  Mereghetti et al.\ (1996) put
upper limits of $\sim 0.1$~mJy for 4.8 GHz radio flux,
$10^{-7}$ photons cm$^{-2}$ s$^{-1}$ for $E > 100$ MeV $\gamma$-rays,
and $V > 25$ for an optical counterpart.
Zavlin, Pavlov \& Tr\"umper (1998)
showed that the {\sl ROSAT} and {\sl ASCA} spectra are consistent with
hydrogen or helium NS atmosphere models.  For a NS of mass $1.4\, M_\odot$
and radius 10 km, they obtained a NS surface temperature
$kT_{\rm eff}^\infty\equiv kT\,(1+z)^{-1}=0.12$--0.16 keV, where $z$ is the
gravitational redshift at the NS surface, and a distance 1.6--3.3 kpc,
compatible with the distance to the SNR,  $d=2.1^{+1.8}_{-0.8}$ kpc
(Giacani et al.\ 2000).

First {\sl Chandra} observation of 1E1207 in January 2000
with a 30 ks exposure allowed us to detect a period $P_1=0.4241296$ s
(frequency $f_1=2.357770$ Hz),
which proved that the source is indeed a NS, with a period typical
for an ordinary radio pulsar (Zavlin et al.\ 2000; Paper I hereafter). 
The pulsar was again observed with {\sl Chandra} for 30 ks in January 2002.
This observation showed a longer period, $P_2=0.4241309$ s ($f_2=2.357763$
Hz), corresponding to the period derivative 
$\dot{P}\approx 2\times 10^{-14}$ s s$^{-1}$
(Pavlov et al.\ 2002b; Paper II hereafter).  Such a period derivative implies
a characteristic age, $\tau_c\equiv P/(2\dot{P}) \sim 300$ kyr, much larger
than the 3--20 kyr age of the SNR (Roger et al.\ 1988), which suggests that
the pulsar was born with a period only slightly faster than its current
value. The conventional magnetic field, $B\sim 3\times 10^{12}$ G,
and the spin-down energy loss rate, $\dot{E}\sim 1\times 10^{34}$ erg s$^{-1}$,
inferred from the $P$, $\dot{P}$ values, are typical for radio pulsars.

Even more surprising finding from the two {\sl Chandra} observations
was the discovery of absorption lines, first lines detected in the spectrum
of an isolated NS (Sanwal et al.\ 2002a).  Two firmly detected lines, 
at 0.7 keV and 1.4 keV, could be interpreted as absorption lines of
once-ionized helium in a very strong magnetic field, 
about $1.5\times 10^{14}$ G,
which requires a gravitational redshift $z=1.12$--1.23 (Sanwal et al.\ 2002a).
Another interpretation, that the lines could be associated with transitions
in He-like oxygen ions in a magnetic field $B\sim 10^{11}$ G, was discussed
by Hailey \& Mori (2002). A possible third line, at about 2 keV (Sanwal et
al. 2002a), unfortunately coincides with the Ir M
line from the telescope mirror coating, where the calibration is inaccurate.
If this line is due to the source, then the three lines might be the fundamental
and two harmonics of the electron cyclotron absorption in a magnetic field
$\sim 0.6\,(1+z)\times 10^{11}$ G. Sanwal et al.\ (2002a) discussed this 
possibility and concluded that it is difficult to reconcile this interpretation
with the much higher magnetic field estimated from $P$ and $\dot{P}$, and
with the very low oscillator strengths of cyclotron harmonics.

First observation of 1E1207 with \xmm\ 
(27 ks exposure) was performed in December 2001.
Mereghetti et al.\ (2002) confirmed the absorption lines and reported 
on their pulse phase variations.  The period value was 
in agreement with that measured in the second {\sl Chandra}
observation, taken 13 days later.

1E1207 was again observed with \xmm\ 
in August 2002 (258 ks of total exposure).
Based on this very deep observation, Bignami et al.\ (2003)
reported positive detection of the 2.1 keV absorption line and marginal
detection of a 2.8 keV line in the source spectrum. The detection of three
(possibly four) evenly spaced lines argues for the above-mentioned cyclotron
interpretation, although it remains unclear how the cyclotron harmonics can
form in the relatively cold plasma with the low magnetic field. Timing analysis
of these data (described below) provides a highly accurate period that
is {\em shorter} than the period measured in the first \xmm\ observation,
indicating that the pulsar is not spinning down steadily.

Finally, {\sl Chandra} observed this puzzling pulsar 
two times in June 2003 for 280 ks of total exposure. These observations
were primarily designed to measure the phase-dependent spectrum
with a high energy resolution. However, they
also provided timing information that we use to assess the evolution of the
pulsar's period. We describe the \xmm\ and {\sl Chandra} observations and 
present results of our timing analysis in \S\,2. 
Possible interpretations are discussed in \S\,3.
The results are summarized in \S\,4.

\section{Observations and timing analysis}

\subsection{\xmm\ observation of December 2001}
The timing analysis of the first (2000 January 6--7) and second 
(2002 January 5--6) {\sl Chandra} observations 
has been described in detail in Papers I and II. 
First timing results for the 
\xmm\ observation of December 2001 have been presented
by Mereghetti et al.\ (2002).
To evaluate the most plausible frequency and its uncertainty
in a statistically rigorous way, and for the sake of uniformity,
we reanalyzed these data with the same approach as in Papers I and II,
using the method of Gregory \& Loredo (1996) based on the Bayesian formalism.
While this method yields frequency estimates consistent with
those given by other techniques (e.\ g., the simple epoch-folding
$\chi^2$ and Rayleigh $Z^2_1$ tests), it has several important advantages.
It is free of any assumption on pulse shape, that results in more accurate
determination of signal frequency and its uncertainty. 
The Bayesian approach implements the phase-averaged
epoch-folding algorithm to calculate the frequency-dependent odds-ratio $O(f)$
that allows one to find the corresponding probability distribution 
$p(f)\propto O(f)$, which in turn is very useful for the interpretation
of the results (see Gregory \& Loredo 1996 and 
Papers I and II for more details).  In the following, we
count all frequencies from a reference value $f_{\rm ref}=2.3577$ Hz. For each
of the epochs, we measure the mean frequency $f_{\rm mean}$ with the
standard deviation $\sigma$ and provide them in the form $\bar{f}=f_{\rm mean}
\pm \sigma$ (e.g., $\bar{f}_1^{\rm ch} = 69.9\pm 1.3$ $\mu$Hz and 
$\bar{f}_2^{\rm ch}=62.5\pm 3.7$ $\mu$Hz
for the first and second {\sl Chandra} observations --- see Paper II). 
We also measure the median
frequency $f_{\rm med}$ with the 68\% and 90\% uncertainties,
$\sigma^{\pm}_{68}$ and $\sigma^\pm_{90}$, below and above $f_{\rm med}$,
and provide them in the form
$f=f_{\rm med}(-\sigma^-_{68},+\sigma^+_{68}; -\sigma^-_{90},+\sigma^+_{90})$   
[e.g., $f_1^{\rm ch} = 69.9\, (-1.4,+1.2;\, -2.5,+1.8)$ $\mu$Hz,
$f_2^{\rm ch} = 62.5\, (-3.2,+4.5;\, -4.2,+8.0)$ $\mu$Hz].

First observation of 1E1207 with \xmm\ was performed on 2001 December 23--24
(orbit 374).  For the timing analysis, we used the data taken with the European
Photon Imaging Camera based on the `p-n' CCD technology (EPIC-pn) operated
in Small Window mode, which provides a $4\farcm4\times 4\farcm4$ sky image and
a 5.7 ms time resolution.  The total observation time span was
27.0 ks, corresponding to an effective exposure of 18.9 ks (because of about
30\% time loss during the CCD readout).  The data were processed with the
most recent version of the ``oal'' library (v.\,3.108)\footnote{To be 
implemented in the SAS-6.0 software (see {\tt http://xmm.vilspa.esa.es/});
the previously known problems with the EPIC-pn timing are fixed in this version
(Kirsch et\ al. 2003).}.  We used 26\,778 counts extracted
from a $30''$-radius circle centered at the source position in the 0.2--5.0
keV energy range.  We estimated that about 88\% of those counts belong to the
source.  The photon arrival times were transformed to the Solar System
Barycenter with the ``barycen'' task.
The $Z_n^2$ test (see Papers I and II for details) gives the most significant
peak, $Z_1^2 = 90.9$, at the frequency of 63.1 $\mu$Hz, very close to 
$63.0\pm 2.6$ $\mu$Hz found by Mereghetti et al.\ (2002) from the 
$\chi^2$ test  assuming that the signal has a sinusoidal shape.
Implementing the Bayesian approach results in the 
probability density distribution $p(f)$ shown in Figure 1. The mean and median
frequencies of the corresponding probability distribution are 
$\bar{f}_1^{\rm xmm}=59.2 \pm 1.3$ $\mu$Hz
and $f_{1}^{\rm xmm} = 59.3\, (-1.4,+0.9;\,  -2.5,+1.5)$ $\mu$Hz.
These values are in agreement
with the frequencies obtained from the $Z_1^2$ and
$\chi^2$ tests.  They do not show statistically significant differences
with the frequency $f_2^{\rm ch}$ (see above) found from the
second {\sl Chandra} observation 13 days later.

\subsection{\xmm\ observations of August 2002}
\xmm\ observed 1E1207 again on 2002 August 4--5
(orbit 486) and 6--7 (orbit 487) with the same instrumental setup
as in the observation on orbit 374.
We excluded intervals of strong background flares 
at the beginning and at the end of each observation and 
used uninterrupted spans of 107.0 and 111.0 ks from the data taken on 
orbits 486 and 487,
respectively (the total effective exposure of EPIC-pn is 152.6 ks). 
The time gap between these two intervals is 73.0 ks.
Applying the same extraction radius and energy range as for the orbit 374
observation, we obtained 108\,945 and 113\,720 
counts from the two data sets for the timing
analysis (background contamination is about 11\%). The frequency distributions
of the odds-ratios 
give mean frequencies $\bar{f}_{2a}^{\rm xmm}=63.31\pm 0.48$
and $\bar{f}_{2b}^{\rm xmm}=64.18\pm 0.33$ $\mu$Hz, and median 
frequencies
$f_{2a}^{\rm xmm} = 63.19\, (-0.26, +0.78;\, -0.58, +0.14)$ $\mu$Hz and
$f_{2b}^{\rm xmm} = 64.30\, (-0.53, +1.08;\, -0.71, +0.39)$ $\mu$Hz
for the orbits 486 and 487, respectively.  These estimates yield
the frequency shift between the two orbits, $\Delta=+0.88\pm0.59$ $\mu$Hz
(calculated from the probability distribution for the frequency
difference --- see Paper II).
The probability that the shift is positive is $P(\Delta>0)=0.932$. 

This shift, albeit of low statistical significance, indicates a positive
frequency derivative, $\dot{f} \sim 10^{-12}$--$10^{-11}$ s$^{-2}$, 
in the joint data set ($T_{\rm span}=291.0$ ks). 
To examine the possible frequency change, 
we calculate the odds-ratio $O(f,\fd)$
and the Rayleigh statistic $Z^2_1(f,\fd)$ 
on a two-dimensional $(f,\,\fd)$ grid,
assuming a linear change of the spin frequency within this observation and
choosing the middle of the total observation time span 
as the reference epoch to minimize the $f$-$\fd$ correlation.
The Rayleigh test gives $Z^2_{1,\rm max}=448.1$ at $f=63.41$ $\mu$Hz,
$\fd=+1.7\times 10^{-12}$ s$^{-2}$. 
To estimate the mean and median parameters and their uncertainties, we used
the probability density distribution $q(f,\fd) = A\, O(f,\fd)$,
where $A$ is the normalization constant, and obtained
$\bar{f}_2^{\rm xmm}=63.49\pm 0.08\ \mu{\rm Hz}$, 
$\bar{\dot{f}}_2^{\rm xmm} = (+2.1\pm 2.0) \times 10^{-12}\,{\rm s}^{-2}$,
\begin{eqnarray}
f_2^{\rm xmm} &  = & 63.51\, (-0.09,+0.06;\, -0.16,+0.10)\,\, \mu{\rm Hz}\,, 
\nonumber \\
\fd_2^{\rm xmm} & = & +2.0\, (-1.8,+2.2;\, -2.4,+2.6) \times 10^{-12}\,\, 
{\rm s}^{-2}\,.
\end{eqnarray}
These estimates suggest that the frequency was indeed increasing during
that observation [$P(\fd_2^{\rm xmm} > 0)=0.908$], but the uncertainties
are too high to conclude this firmly.  The probability density
distribution for frequency, $p(f)=\int q(f,\fd)\, {\rm d}\fd$,
is shown in Figure 1.

The probability distribution for the frequency difference 
between the \xmm\ observations of August 2002 and December 2001 
is shown in the upper panel of Figure 2.  
The mean and median values for the difference
are $\bar{\Delta}=+4.30\pm 1.28$ $\mu$Hz and
$\Delta= +4.26$\,($-$0.81, +1.08; $-$1.69, +2.31) $\mu$Hz.
The probability that the pulsar has {\em spun up} in about 7.5 months
since the first \xmm\ observation is $P(\Delta >0)=0.992$.
If the frequency were increasing monotonously during this period 
($1.9\times 10^7$ s), then its derivative would be in the range
$\fd=(1.4-3.4)\times 10^{-13}$ s$^{-2}$ (at a 90\% confidence level).

\subsection{Chandra observations of June 2003}
1E1207 was observed with 
the Low Energy Transmission Grating in combination
with the Advanced CCD Imaging Spectrometer operated in
Continuous Clocking mode 
on 2003 June 10--12 (155.7 ks) and June 18--19 (115.2 ks). 
This observational mode provides a 2.9 ms time resolution
by means of sacrificing spatial resolution in one dimension.
For the timing analysis, we used 11\,909 and 8\,804 counts 
(for the first and second data sets, respectively) extracted 
from segments of a 7-pixel (3\farcs44) 
width in the zero-order images, in the energy range of 0.2--5.0 keV.
The times of arrival were corrected for the dither and the Science Instrument
Module motion as described in Paper I and transformed to the Solar System
Barycenter using the ``axBary'' tool of the CIAO package\footnote{
{\tt http://asc.harvard.edu/ciao/}}.

The mean and median frequencies, as given by odds-ratios, 
are $\bar{f}_{3a}^{\rm ch}=64.11 \pm 0.96$, 
$\bar{f}_{3b}^{\rm ch}=62.47 \pm 1.36$ $\mu$Hz,
and $f_{3a}^{\rm ch}=64.35$ (--1.40, +0.64; --1.93, +1.00) $\mu$Hz,
$f_{3b}^{\rm ch}=62.09$ (--0.59, +1.78; --1.04, +3.23) $\mu$Hz
(the subscripts `3a' and `3b' are 
related to the observations performed on June 10--12 and 18--19, respectively).
The frequencies of the $Z_1^2$ peaks, 64.0 $\mu$Hz ($Z_{\rm 1,max}^2=27.5$)
and 63.5 $\mu$Hz ($Z_{\rm 1,max}^2 = 25.4$), are within the uncertainties of
the mean and median frequencies. 
The mean and median values of the frequency difference
between the two observations are
$\bar{\Delta} = -1.67\pm1.62$ $\mu$Hz
and $\Delta = -1.98$\, (--1.10, +1.84) $\mu$Hz.
The probability that the pulsar has {\em spun down} in a week
between the two observations is $P(\Delta < 0) = 0.853$.

We also performed the timing analysis of the combined data set 
($T_{\rm span}=799.9$ ks) on a two-dimensional grid of $f$ and $\fd$, 
using the same approach as in \S\,2.2. The maximum value
of the Rayleigh statistic is
$Z^2_{1,{\rm max}}=52.7$, at $f=63.61$ $\mu$Hz,
$\fd=-1.0\times 10^{-12}$ s$^{-2}$.
Because of the large time gap, $T_{\rm gap}=529.1$ ks, the phase
coherence between the two data sets was lost, 
that resulted in a number of peaks (frequency aliases) in the
frequency dependences of $O$ and $Z_1^2$,
separated by $\approx 1.4$--1.5 $\mu$Hz.
Three most significant peaks (we denote them as A, B, and C)
are seen in the probability density distribution $p(f)$ 
in the bottom panel of Figure 1.  The mean and median 
parameters as estimated
from the the multi-peak two-dimensional distribution are
$\bar{f}_3^{\rm ch} = 63.37\pm 0.74$ $\mu$Hz, 
$\bar{\fd}_3^{\rm ch}=(-3.0\pm 2.1)\, \times 10^{-12}\, {\rm s}^{-2}$, and
$f_3^{\rm ch} = 63.58\, (-0.10,+0.07;\, -1.37,+0.42)$ $\mu$Hz,
$\fd_3^{\rm ch} = -3.5\, (-1.6,+2.5;\, -2.7,+3.9)\times 10^{-12}\, 
{\rm s}^{-2}$.
Considering $(f,\, \fd)$ domains around the peaks A, B and C separately
results in the following mean and median frequencies:
$\bar{f}_A= 62.10\pm 0.05\,\, \mu{\rm Hz}$, 
$\bar{f}_B= 63.59\pm 0.05\,\, \mu{\rm Hz}$,
$\bar{f}_C= 64.96\pm 0.07\,\, \mu{\rm Hz}$,
\begin{eqnarray}
 f_A & = & 62.09\, (-0.03,+0.04;\, -0.08,+0.10)\,\, \mu{\rm Hz}\, , \nonumber \\
 f_B & = & 63.60\, (-0.06,+0.04;\, -0.16,+0.11)\,\, \mu{\rm Hz}\, , \nonumber \\
 f_C & = & 65.02\, (-0.06,+0.09;\, -0.12,+0.17)\,\, \mu{\rm Hz}\, , 
\end{eqnarray}
and frequency derivatives:
$\bar{\fd}_{A,-12} = -1.6 \pm 1.4$,
$\bar{\fd}_{B,-12}=-3.6\pm 1.8$,
$\bar{\fd}_{C,-12}=+0.4\pm 1.6$,
\begin{eqnarray}
 \fd_{A,-12} & = & -1.4\,  (-1.2,+0.9;\, -2.0,+1.7)\, , \nonumber \\ 
 \fd_{B,-12} & = & -3.8\,  (-1.2,+1.6;\, -1.8,+2.6)\, , \nonumber \\
 \fd_{C,-12} & = & +0.5\,  (-1.5,+1.1;\, -2.5,+1.7)\, ,
\end{eqnarray}
where $\fd_{-12}=\fd/(10^{-12}\, {\rm s}^{-2})$.
The relative contributions of the peaks A, B, C into the probability 
distribution (i.e., the probabilities that the true frequency 
and its derivative are associated with a
given peak) are $P_A=0.207$, $P_B=0.734$, $P_C=0.059$. 
Therefore, although the parameters related to peak B are
more probable, none of the two other parameter sets can be 
ruled out on statistical grounds.

The probability distribution of the frequency difference between June 2003
and  August 2002, plotted in the lower panel of
Figure 2, also has 3 discernible peaks.  
The mean frequency shift in the 10 month period is only 
$\bar{\Delta}= +0.06\pm 0.61$ $\mu$Hz;
the median frequency difference is
$\Delta=+0.09$\, (--0.12,+0.10;\, --1.46, +0.14) $\mu$Hz.
If we assume that peak A (or B or C)
corresponds to the true frequency in June 2003, then the shifts are 
$\bar{\Delta}_A = -1.36 \pm 0.10$ $\mu$Hz and
$\Delta_A=-1.37$\, (--0.07,+0.09;\, --0.10,+0.16) $\mu$Hz,
or $\bar{\Delta}_B = +0.10 \pm 0.07$ $\mu$Hz and
$\Delta_B=+0.11$\, (--0.05,+0.07;\, --0.09,+0.12) $\mu$Hz,
or $\bar{\Delta}_C = +1.47 \pm 0.09$ $\mu$Hz and
$\Delta_C=+1.48$\, (--0.08,+0.07;\, --0.14,+0.18) $\mu$Hz.
Assuming a linear time dependence of frequency during
this period ($2.7\times 10^7$ s), we can constrain the frequency derivative,
$-5.5\times 10^{-14} < \fd < +6.0\times 10^{-14}$ s$^{-2}$, 
at a 90\% confidence level.

\section{Discussion}
It is easy to check that the time dependence of 
the pulsation frequency in the 3.45 yr 
interval cannot be satisfactorily fitted by a straight line, 
$f(t)=f_0 + \dot{f} (t-t_0)$,
for any of the three possible frequency values obtained in the third
{\sl Chandra} observation.  The best fit, 
with $f_0=68.0$ $\mu$Hz, $\dot{f} = -5.2\times 10^{-14}$ s$^{-2}$
($t_0=51\,500.0$ MJD), is obtained assuming the correct frequency 
in June 2003 is given by peak A (see the dotted line in 
Fig.\ 3); the fit corresponds to
$\chi^2_{\rm min}=22.6$ (for 3 degrees of freedom) and can be
rejected at a 4.1$\sigma$ level (if one assumes the frequency of peak B
for the third {\sl Chandra} observation, then the rejection level is
5.6$\sigma$).
This means that 1E1207 is not spinning down steadily, as most radio pulsars do.
We discuss possible explanations below.

\subsection{A glitching pulsar?}
Some radio pulsars and AXPs occasionally show sudden increases in pulsation
frequency, commonly known as ``glitches'', with various patterns of post-glitch
behavior  (e.g., Lyne, Shemar \& Graham-Smith 2000; Gavriil \& Kaspi 2002). 
The relative frequency jumps, $\Delta f/f$, vary from $\sim 10^{-9}$ to 
$\sim 10^{-5}$ in different pulsars. Obviously, the non-monotonous behavior of
$f(t)$, inferred from the above-described timing of 1E1207, may suggest that
the pulsar experienced a number of glitches during the 3.45 yr time span.
Assuming the glitches are not associated with a substantial change of frequency 
derivative $\fd$, at least two glitches are required, 
between January and August 2002 (212 days interval), and between August 2002
and June 2003 (313 days interval).  The values of the frequency jumps depend on
the value of $\fd$ assumed, which in turn depends on the cumulative
frequency change in glitches that might occur in the 716 days interval
between January 2000 and December 2001. For instance, if there were no glitches
in that interval, then $\fd \simeq -1.6\times 10^{-13}$ s$^{-2}$,
and the two frequency jumps, of about 7 $\mu$Hz and 5 $\mu$Hz,
are required to fit the data  
(the latter value assumes that the correct frequency in 2003 June 10--19 
is given by the central peak B of the probability distribution --- see Fig.\ 3).
The larger of these jumps ($\Delta f/f \sim 3\times 10^{-6}$), is similar to
the strong glitches observed in the Vela pulsar (Dodson,
McGulloch \& Lewis 2002) and the
AXP 1E~2259+586 (Kaspi et al.\ 2003), and it is a factor of 5 smaller than the
giant glitch recently observed in PSR J1806$-$2125 (Hobbs et al.\ 2002).

Given that at least two glitches are required in the 525 
day interval, it
seems reasonable to assume that the pulsar was also glitching in the 717 
day interval between the first two observations, which would correspond to a
larger $|\fd|$ and stronger glitches. For instance, if the cumulative
frequency increase due to glitches was 30 $\mu$Hz in 717 days, then 
$\fd \simeq -6.6\times 10^{-13}$ s$^{-2}$, and the two 
frequency jumps (in 525 days after January 2002) are about 16 $\mu$Hz
and 18 $\mu$Hz ($\Delta f/f \sim 7\times 10^{-6}$). Although the relative
frequency jumps of the separate glitches are smaller than for the largest
glitch observed (Hobbs et al.\ 2002), the integrated amplitude of the glitches
is a factor of 30--60 greater then the typical value,
$\sum\Delta f \sim 0.02 |\fd| T$ ($T$ is the total time span,
equal to 3.45 yr in
our case) observed for glitching radio pulsars (Lyne et al.\ 2000).
Therefore, although we cannot formally rule out the possibility that 1E1207
experiences glitches of $\Delta f \gtrsim 5\, \mu$Hz, with a characteristic
rate of $\sim$0.5--1 
glitches per year, such a hypothesis implies that the nature
of the glitches in this unusual object is different from 
that observed in radio pulsars.
In addition, to explain the probable increase of frequency between
the two orbits in the \xmm\ observation of August 2002 (see \S2.2),
we have to assume that a small glitch of $\sim 1\, \mu$Hz occurred
in the short 73 ks gap, which looks artificial.
It is worth mentioning that the glitch hypothesis implies a smaller
characteristic age (e.g., $\tau \sim 60$ kyr in the last example), bringing it
closer to the SNR age, and larger magnetic field ($B\propto |\fd|^{1/2}$)
and spin-down energy loss ($\dot{E}\propto |\fd|$).

\subsection{Accretion as the source of the spin frequency variations?}
The spin evolution of a NS can also be affected by a flow 
of material streaming to the NS from an accretion disk.
The analysis presented in \S3.3 below rules out an accreting close binary,
but the accreting material could be supplied from a ``fossil'' accretion disk 
formed from ejecta produced in the supernova explosion (see Marsden,
Lingenfelter \& Rotschild 2001 for references).
The accretion can proceed in two regimes (e.g., Frank, King, \& Raine 2002), 
depending on the relation between the corotation radius
\begin{equation}
r_c = (G\,M)^{1/3} (2\pi f)^{-2/3} = 0.1\times 10^9\,\,{\rm cm},
\end{equation}
(here $f\simeq 2.36$ Hz is the spin frequency)
and magnetospheric radius 
\begin{equation}
r_m \sim 0.5 (8\,G\,M)^{-1/7} \mu^{4/7} \dot{m}^{-2/7}
\simeq 0.9\times 10^9
\dot{m}_{14}^{-2/7}\,\mu_{30}^{4/7}\,\, {\rm cm}
\end{equation}
(here $\dot{m}=10^{14}\,\dot{m}_{14}$ g s$^{-1}$ is the accretion rate,
$\mu=10^{30}\, \mu_{30}$ G cm$^3$ is the NS magnetic moment,
and the NS mass is assumed to be $M=1.4\,M_\odot$).
If the accretion rate is so high that $r_m<r_c$, the accreting matter can
overcome the centrifugal barrier and reach the NS surface.
In this ``accretor regime'' the torque exerted on the magnetosphere    
spins the NS up ($\fd>0$).
At $r_m>r_c$, the centrifugal force at $r=r_m$ exceeds
the gravitational force,
so that accretion onto the NS surface is inhibited.
In this ``propeller regime'' the infalling material is accelerated away
from the magnetosphere reducing the angular momentum of the NS
($\fd < 0$).  At even lower $\dot{m}$, $r_m$ approaches the
light cylinder radius, $r_{\rm lc}=c/(2\pi f)=2.0\times 10^9$ cm,
where the pressure of the magneto-dipole radiation of the rotating NS
takes over the accretion pressure and prevents any accretion (``ejector
regime''). 

The torque caused by accretion in both the propeller and accretor regimes
can be conveniently approximated, at $r_m\ll r_{\rm lc}$, as
\begin{equation}
K\approx 4\pi\,\dot{m}\,r_m^2\, (f_{\rm eq} - f)
\end{equation}
(e.g., Menou et al.\ 1999), where 
$f_{\rm eq} = (2\pi)^{-1}\,(G M)^{1/2}\,r_m^{-3/2} = 
8.0\times 10^{-2}\, \dot{m}_{14}^{3/7}\mu_{30}^{-6/7}$ Hz is the spin
frequency at which the propeller-accretor transition occurs 
($f=f_{\rm eq}$ at $r_m=r_c$). 
The corresponding frequency derivative, $\fd = K (2\pi I)^{-1}$, is
\begin{equation}
\fd \approx 1.6\times 10^{-13}\,
\dot{m}_{14}^{3/7} \mu_{30}^{8/7}\,(f_{\rm eq} -f)\,\,{\rm s}^{-2},
\end{equation}
for the NS moment of inertia $I=10^{45}$ g cm$^2$. 
The curves of constant $\fd$ values in different regimes are shown in Figure 4.

If the observed variations of $\fd$ are due to accretion,
the accretion rate should vary with time. Moreover,
to explain the very plausible {\em spin-up}
between December 2001 and August 2002 (see \S2.2),
we have to assume, in the framework of this simple model,
that 1E1207 was in the accretor regime during at least part of that time
interval.  Figure 4 shows that in order to provide 
the corresponding $\fd \gtrsim +10^{-13}$ s$^{-2}$,
the accretion rate should be high enough, 
$\dot{m}_a\gtrsim 10^{16}$ g s$^{-1}$
($\dot{m}_a$ [$\leq \dot{m}$] is the rate of accretion onto the NS surface).
Because accretion onto the NS surface provides an additional luminosity
$\Delta L \sim G M \dot{m}_a R^{-1} \simeq 1.9\times 10^{36}\, 
\dot{m}_{a,16}$ erg s$^{-1}$,
which substantially exceeds the observed X-ray luminosity, 
$L_X\simeq  1.5\times 10^{33} d_{2.1}^2$ erg s$^{-1}$ in the 0.3--5 keV range 
($d_{2.1} =d/[2.1\,{\rm kpc}]$),
 we have to assume that 1E1207 was in the propeller
stage during all the available X-ray observations, while it 
perhaps was in the accretion stage (accompanied by 
an unnoticed outburst) between January and August of 2002. 
(This outburst plays the same role as the 7 $\mu$Hz glitch assumed
in the glitching interpretation.)

Assuming no other accretion outbursts occurred in the 3.45 yr of the
{\sl Chandra} and \xmm\ observations, typical spin-down rates are in the
range of $|\fd|\sim 10^{-14}$--$10^{-13}$ s$^{-2}$. 
According to Figure 4, in the propeller regime
they correspond to $\dot{m}\sim 10^{12}$--$10^{16}$ g s$^{-1}$
and $\mu\sim 10^{29}$--$10^{30}$ G cm$^3$ (or $B\sim 10^{11}$--$10^{12}$ G
for a centered magnetic dipole).
The higher values of $\dot{m}$ from the above
range are hardly plausible because such strong accretion onto
the magnetosphere should be accompanied by emission
(presumably, optically thin bremsstrahlung -- see Wang \& Robertson 1985)
with a luminosity $L\sim GM\dot{m}r_m^{-1}\sim 10^{34}\,
\dot{m}_{16}^{9/7}\mu_{30}^{-4/7}$ erg s$^{-1}$.
The lack of such emission in the observed data means that 
$\dot{m}\lesssim 10^{14}$--$10^{15}$ g s$^{-1}$. 
At lower $\dot{m}$ ($\lesssim 10^{13}$ g s$^{-1}$),
the X-ray pulsar approaches the propeller-ejector boundary,
where the propeller spin-down becomes less efficient but the radio pulsar can
turn on. If $\dot{m}$ varies around this boundary, 1E1207 can occasionally
manifest radio-pulsar activity.   

In the propeller regime, we can expect variations of the accretion
rate comparable with its average value, to explain the variations
of $\fd$. If a fraction of $\dot{m}$ accretes onto the NS surface
(which is possible even in the propeller regime 
--- e.g., Menou et al.\ 1999),
it could contribute to the observed X-ray luminosity
and lead to its variations.
The fact that the observations of 1E1207 at different 
epochs do not reveal significant
changes in its X-ray flux allows one to constrain this fraction.
For instance, the difference between the fluxes derived from the two 
\xmm\ observations does not exceed 1\%, which corresponds to 
$\Delta L_X < 1.5\times 10^{31} d_{2.1}^2$
erg s$^{-1}$ and $\Delta\dot{m}_a \sim \dot{m}_a < 10^{11}\,\, 
{\rm g}\,\, {\rm s}^{-1} \ll \dot{m}$. Observations of 1E1207 with different
X-ray observatories show flux differences of up to 30\%, but 
they can likely be explained by systematic uncertainties in 
instrument calibrations.

Strong constraints on the properties of the putative accretion disk
can be obtained from optical observations of 1E1207.
Using a standard approach (e.g., Frank et al.\ 2002), we can estimate the 
effective temperature of a geometrically thin, optically thick disk as
\begin{equation}
T(r) = (3GM\dot{m})^{1/4}(8\pi r^3\sigma)^{-1/4} = 
T_{\rm in} (r_{\rm in}/r)^{3/4}\,,
\end{equation}
where 
\begin{equation}
T_{\rm in} =1.5\times 10^4\, \dot{m}_{14}^{13/28} \mu_{30}^{-3/7}\,{\rm K}
\end{equation}
is the temperature at the inner edge of disk, at 
$r_{\rm in}\approx r_m$.  The spectrum
of such a disk, in the blackbody approximation, is
\begin{equation}
F_\nu = \frac{4\pi h \nu^3\,\cos i}{c^2 d^2}\int_{r_{\rm in}}^{r_{\rm out}}
\frac{r\,{\rm d}r}{\exp[h\nu/kT(r)]-1}\,,
\end{equation}
where $i$ is the disk inclination.
We calculated the spectral flux $F_\nu$ and the
corresponding $V$ magnitude on a $\mu$-$\dot{m}$ grid and plotted the
lines of constant $V$ in Figure 4 for $\cos i=1$, $d=2.1$ kpc, 
adding a plausible extinction $A_V=0.7$. The results 
are insensitive to the value of the outer radius $r_{\rm out}$ when
it exceeds $\sim 10^{11}$ cm (we used $r_{\rm out}=10^{14}$ cm in
this calculation).

Figure 4 shows that, for the face-on disk,
expected magnitudes in the accretor and propeller regimes are  
$V\sim 17$--19 and $V\sim 20$--23, respectively.
To reconcile the magnitude estimates with the 
reported limit, $V>25$, we have to assume 
inclinations close to $90^\circ$, which looks rather artificial.
On the other hand, the inclination should not be too close to $90^\circ$
in order the disk periphery not to obscure the X-ray source:
$\cos i > H(r_{\rm out})/r_{\rm out} 
\sim 10^{-2}\,\dot{m}_{14}^{3/20} r_{\rm out,12}^{1/8}$,
where $H(r)\propto r^{9/8}$ is the vertical scale-height of the disk, 
in the thin disk model.

To conclude, the accretion hypothesis can, in principle, explain the
observed variations of pulsation frequency and give some constraints
on the magnetic field, but it implies that 1E1207 is a transient X-ray
source, which has not been observed, and requires extreme inclinations
of the accretion disk. 

\subsection{A binary with a low-mass companion?}
Another plausible explanation for the observed frequency variation is that
1E1207 resides in a binary system. 
The orbital motion would result in a periodic
modulation of the observed frequency due to the Doppler shift. We can put
some limits on the nature of the binary system and the companion of 1E1207
from the available X-ray and optical observations.
 
Using our timing results, we can constrain
the amplitude $f_a$ of the orbital modulation
and the orbital period $P_{\rm orb}$.  The observed 
deviations from a steady spin-down imply $2 \lesssim f_a \lesssim 20$ $\mu$Hz
(larger $f_a$ values cannot be formally excluded but seem hardly plausible,
given the range of the measured frequencies).
To see a significant deviation from the steady spin-down
over the 3.45 yr time span,
the orbital period should not be larger than about 6
years, for $f_a < 20$ $\mu$Hz. On the other hand,
the orbital period should substantially exceed the time spans of the long
observations of August 2002 ($T_{\rm span}=3.4$ d) and June 2003
($T_{\rm span}=9.3$ d) that do not show frequency variations except
for the linear term, $\fd t$.
To estimate a lower limit on $P_{\rm orb}$, we repeated the timing analysis
of these two observations
for a large number of models with sinusoidally modulated $f(t)$.
The lower limit, naturally, grows with increasing $f_a$, and it strongly
depends on the assumed orbital phase. For example, we obtained
$P_{\rm orb}>(2-50)\,T_{\rm span}$ for $f_a=2$ $\mu$Hz.
This gives the lowest possible period of $\approx 20$ d, for 
a special choice of orbital phase, while the lower limit is 
significantly larger, $P_{\rm orb}\gtrsim 50$--100 d,
for a broad range of orbital phases. 
We also attempted to constrain the lower limit on
$P_{\rm orb}$ fitting the time dependence
\begin{equation}
f(t)=f_0+\fd_0\,t+f_a\,\sin[2\pi\,(t-t_p)/P_{\rm orb}]
\end{equation}
(which assumes a circular orbit) to all five timing points 
shown in Figure 3 and requiring the model to be 
consistent with the timing data in each individual data set.
This resulted in a `global lower limit' $P_{\rm orb}\gtrsim 60$ d.

Within the possible period boundaries, multiple combinations of 
orbital parameters can give ``perfect fits'' to the scarce timing
observations.  We present three representative solutions in Figure 5.
These solutions correspond the following parameters in equation (11):
$f_0=65.4$ $\mu$Hz, $\fd_0=-1.8\times 10^{-14}$ s$^{-2}$,
$f_a=4.8$ $\mu$Hz, $P_{\rm orb}=201.4$ d and $t_p=-12.5$ d (Model I),
$f_0=69.3$ $\mu$Hz, $\fd_0=-8.0\times 10^{-14}$ s$^{-2}$,
$f_a=4.6$ $\mu$Hz, $P_{\rm orb}=937.8$ d and $t_p=18.8$ d (Model II),
and $f_0=65.6$ $\mu$Hz, $\fd_0=-1.6\times 10^{-14}$ s$^{-2}$,
$f_a=-6.0$ $\mu$Hz, $P_{\rm orb}=595.2$ d and $t_p=111.0$ d (Model III),
with time $t$ counted from MJD 51\,500.0.
Model I assumes that the correct frequency in the observation
of June 2003 is given by peak A (see \S\,2.3 and Fig.\ 1), whereas
the more probable frequency of peak B is chosen for the two other models. 
We note that in these examples
the characteristic pulsar age, $\tau_c=-f/(2\fd)$,
ranges from 466 to 2\,335 kyr; it is still significantly
larger than the 20-kyr upper limit on the age of the SNR. 
To reconcile the ages in the framework of this simple model, 
one would need a very large, and rather implausible,
orbital modulation $f_a\gtrsim 120$ $\mu$Hz.

The above estimates show that 1E1207 could be a wide binary 
with a long period, $P_{\rm orb}\sim 0.2$--6 yr,
and the component separation,
$a\sim (0.3-3)\,(m_1+m_2)^{1/3}$ AU,
where $M_1=m_1 M_\odot$ and $M_2=m_2 M_\odot$ 
are masses of the NS and its companion, respectively. 
The plausible amplitudes of $f_a$ correspond to 
rather small amplitudes of radial velocity,
$v_{1r}= 0.64\,(f_a/5\,\mu{\rm Hz})$ km s$^{-1}$.
The mass function of the binary can be estimated as
$(m_2\,\sin i)^3\, (m_1 + m_2)^{-2} = 
(P_{\rm orb}/2\pi G)\,v_{1r}^3\,M_\odot^{-1} =
1.0\times 10^{-5} (P_{\rm orb}/1\,{\rm yr})(f_a/5\,\mu{\rm Hz})^3$.
For $\sin i \gg 0.03$, this gives
$m_2\,\sin i = 0.027\,
(P_{\rm orb}/1\,{\rm yr})^{1/3}\,(f_a/5\,\mu{\rm Hz})\,(m_1/1.4)^{2/3}$.
We see that for reasonably large inclinations
the secondary star should have
a very low mass, in the range of brown dwarf or M-dwarf masses,
with a lower limit as small as 6 Jupiter masses.
The upper limit on $m_2$ can be estimated from optical observations.  For
instance, the limiting magnitude $V>25$ (Mereghetti et al.\ 1996) implies the
absolute $V$ magnitude 
$M_V>12.7$ (for $d=2.1$ kpc and a plausible extinction $A_V=0.7$),
corresponding to an M dwarf later than 
M3--M4 (mass $m_2 \lesssim 0.25$--$0.3$)
or a white dwarf (of any allowed white dwarf mass, $m_2\lesssim 1.4$).

Thus, we conclude
that the timing results can be explained by the binary hypothesis
if either the mass of the secondary is very low or 
the inclination is very small.  Similar to the case of glitches, 
the intrinsic $\fd$ (hence the characteristic age, spin-down energy loss,
and magnetic field) can be quite different from the values
inferred in Paper II from two {\sl Chandra} observations.  It
should be noted that a low mass of the secondary implies that
the pre-supernova binary 
had a very eccentric orbit, with the secondary close to
aphelion at the moment of explosion, in order for the binary to survive.
Therefore, such low-mass binaries in SNRs should be a very rare case.
The only potential low-mass binary with a young NS is the central object
of the SNR RCW 103, but likely it is
a short-period ($P_{\rm orb} \approx 6.4$ hr), 
accreting binary (Sanwal et al.\ 2002b).  We note also that if 1E1207 is
indeed in a wide binary system, 
we would not see manifestations of binarity other than the timing properties.

\section{Conclusions}
Our timing analysis of the {\sl Chandra} and \xmm\ observations, 
spread over a 3.45 yr time span, has shown
a non-monotonous frequency evolution of 1E1207. To explain the observed
deviations from the steady spin-down, we have discussed three hypotheses.
The first one, that the neutron star is a glitching pulsator, requires
an unusually large time-integrated amplitude of glitches 
to explain the observed variations in the spin frequency.
The second hypothesis assumes variable
accretion from a dim fossil (residual) disk around
the pulsar and implies that 1E1207 is a transient X-ray source
that can occasionally increase its brightness by at least three
orders of magnitudes, which has never been observed.
The deep limit on optical emission from 1E1207
implies that such a disk is seen almost edge-on.
The third hypothesis is that 1E1207 resides in a wide binary system
with a low-mass companion that has not yet been detected. 
A binary with such a companion could survive the supernova explosion
only at very special conditions.  Thus, all three hypotheses imply rather
exotic properties of 1E1207, but none of them can be firmly ruled out.
We consider the binary hypothesis as somewhat more plausible than
the other two, but only further observations can tell us which
(if any) of these interpretations is correct. The most direct way to
understand the true nature of 1E1207 is to monitor its timing
behavior in a dedicated {\sl Chandra} or \xmm\ program. 
Additional useful constraints on the nature of the
putative low-mass companion or accretion disk can be obtained
from extremely deep optical/IR observations. Finally, it would be
worthwhile to carry out a series of very deep radio observations
to look for a possible transient radio pulsar.

\acknowledgements
We are thankful to Michael Freyberg and Uwe Lammers for their help with issues
concerning the EPIC-pn timing. 
We also thank the anonymous referee for useful remarks. 
Support for this work was provided by the NASA
through Chandra Award GO3-4091X, issued by the Chandra X-Ray Observatory Center,
which is operated by the Smithsonian Astrophysical Observatory for and on behalf
of NASA under contract NAS8-39073. This work was also partially supported by
NASA grant NAG5-10865.


\newpage
\begin{figure}
\plotone{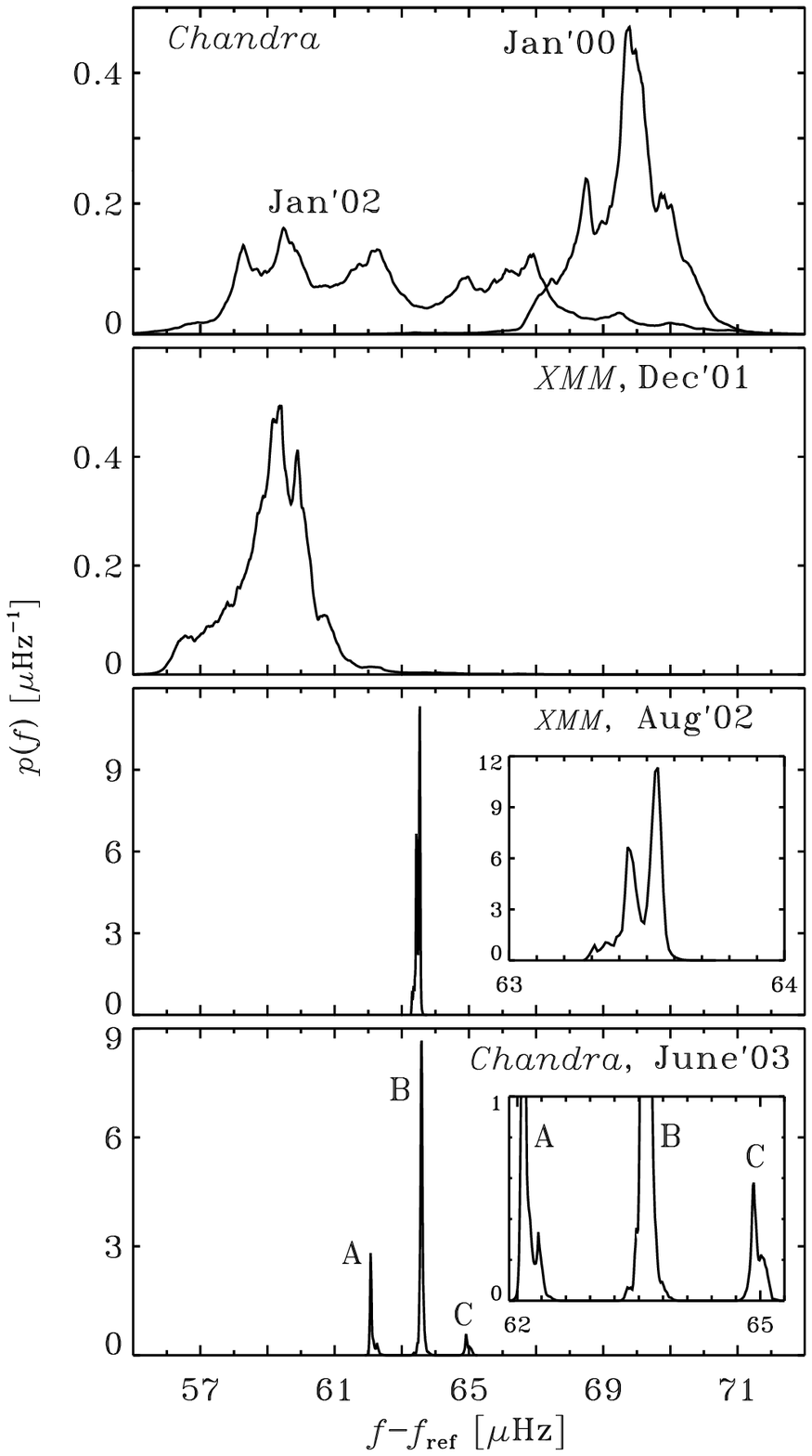}
\caption{
Probability density distributions $p(f)$
for five observations of 1E1207
($f_{\rm ref} = 2.3577$ Hz is the reference frequency).
The insets in the two lower panels present the $p(f)$
dependences in narrower frequency ranges.
}
\end{figure}
\clearpage

\begin{figure}
\plotone{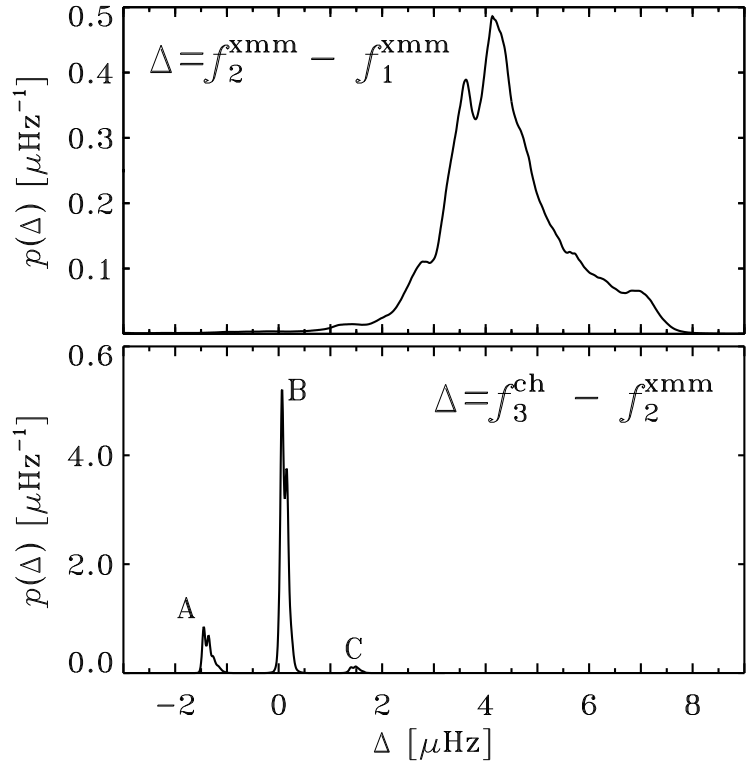}
\caption{Probability distributions $p(\Delta)$
for frequency difference $\Delta$ for two pairs of observations:
August 2002 -- December 2001 (upper panel) and 
June 2003 -- August 2002 (lower panel).
}
\end{figure}
\clearpage

\begin{figure}
\plotone{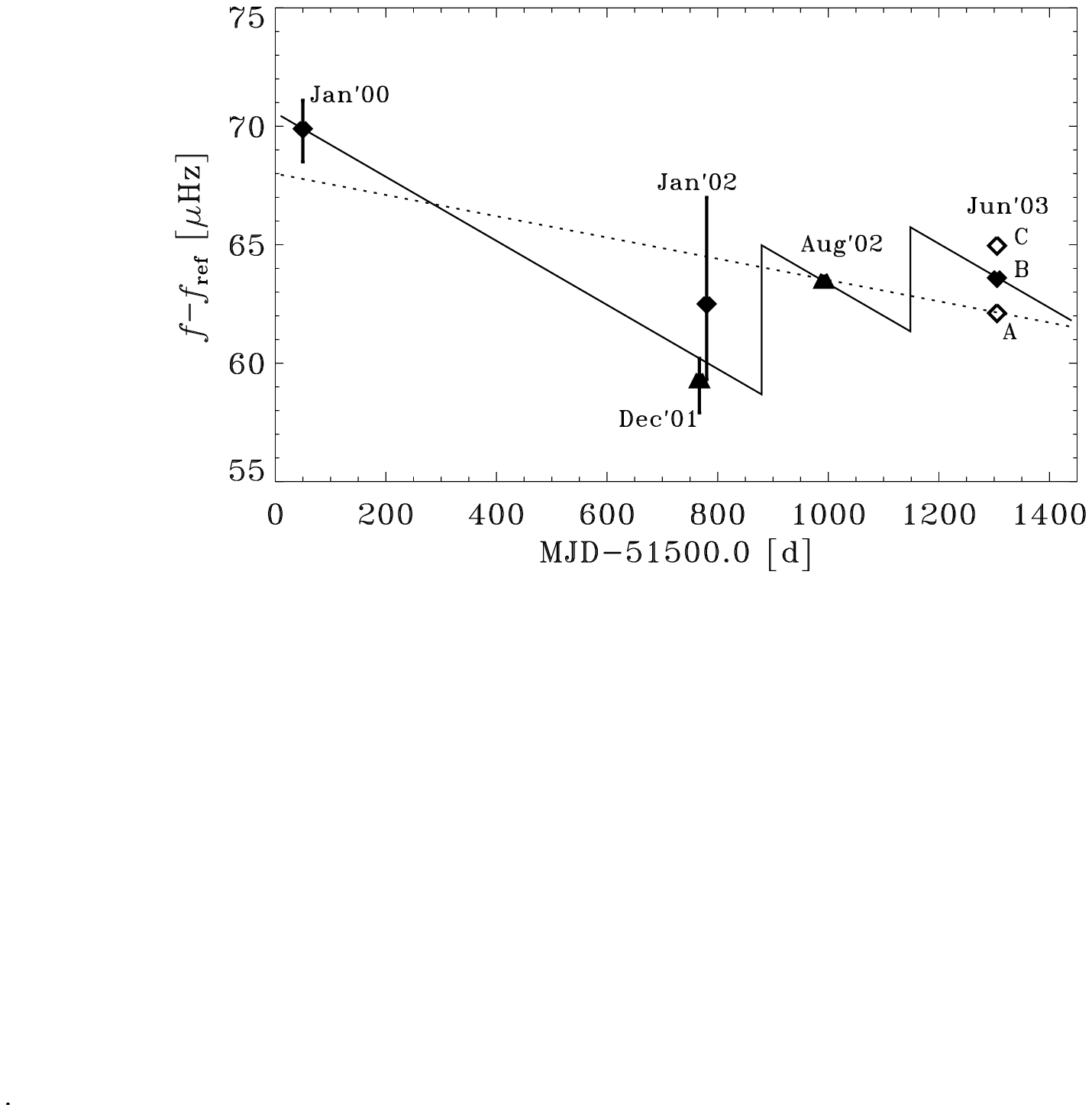}

\caption{
Possible glitching scenario to explain the
observed variations of the pulsation frequency 
(median frequencies with their 68\% uncertainties are plotted). 
The diamonds and triangles indicate the {\sl Chandra} and \xmm\ observations,
respectively.  The points A, B, and C are three timing solutions for
the June 2003 observation (see Fig.\ 1). The solid line 
corresponds to $\fd=-1.6\times 10^{-13}$ s$^{-2}$
between the glitches, and it assumes that
the correct frequency in June 2003 is given by point B.
The dots show the best straight-line fit 
with $\fd=-5.2\times 10^{-14}$ s$^{-2}$, assuming the correct frequency
in June 2003 is given by point A.
}
\end{figure} 
\clearpage

\begin{figure}
\plotone{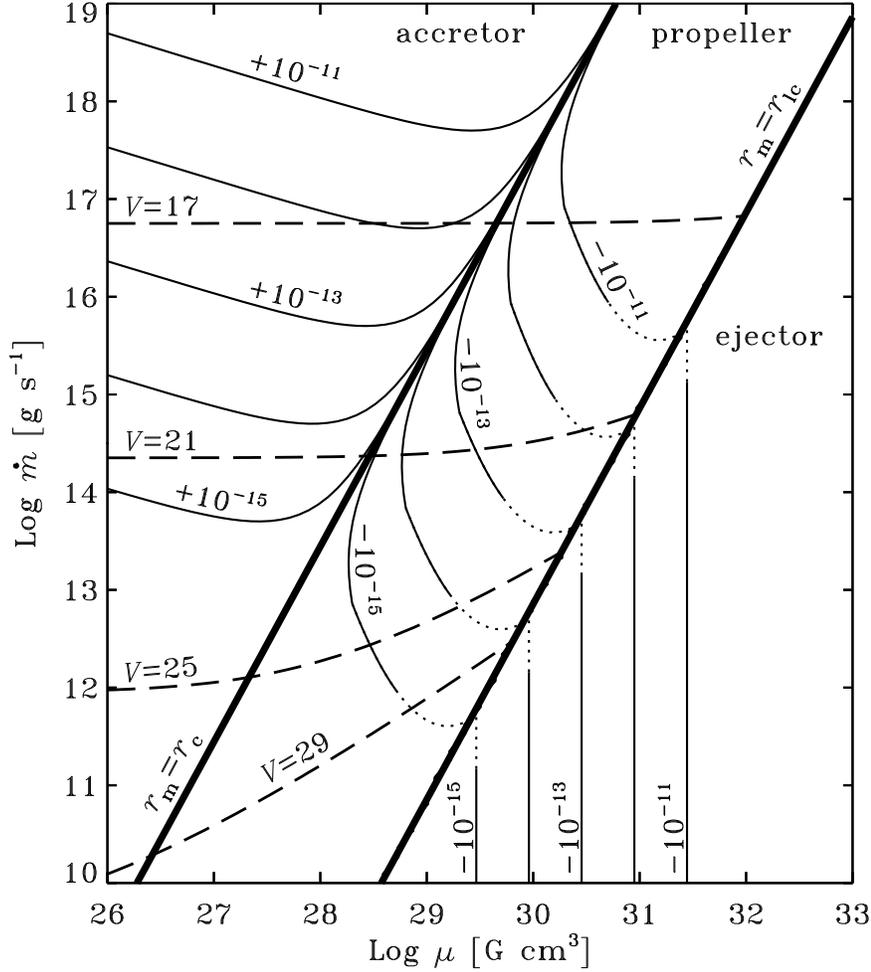}
\caption{
Three regimes of interaction of a NS with accreting matter,
for the NS spin frequency $f=2.36$ Hz (see \S\,3.2). 
The thin solid curves are lines of constant $\fd$ values
(depicted near the curves, in units of s$^{-2}$).
They are obtained from equation (7) for the accretor and propeller regimes.
In the ejector (radio-pulsar) regime, the frequency derivative
is calculated as $\fd = -(3\pi I)^{-1} \mu^2 r_{\rm lc}^{-3}$,
assuming a magneto-dipole braking.  The dotted parts of the curves
in the propeller regime are drawn arbitrarily to demonstrate decreasing
efficiency of propeller braking when the magnetosphere radius $r_m$
approaches the light-cylinder radius $r_{\rm lc}$.
The long-dash curves correspond to constant $V$ magnitudes
of the predicted optical emission from an accretion disk (eq.\ [10]),
for $A_V=0.7$ and $i=0$. 
}
\end{figure}

\begin{figure}
\plotone{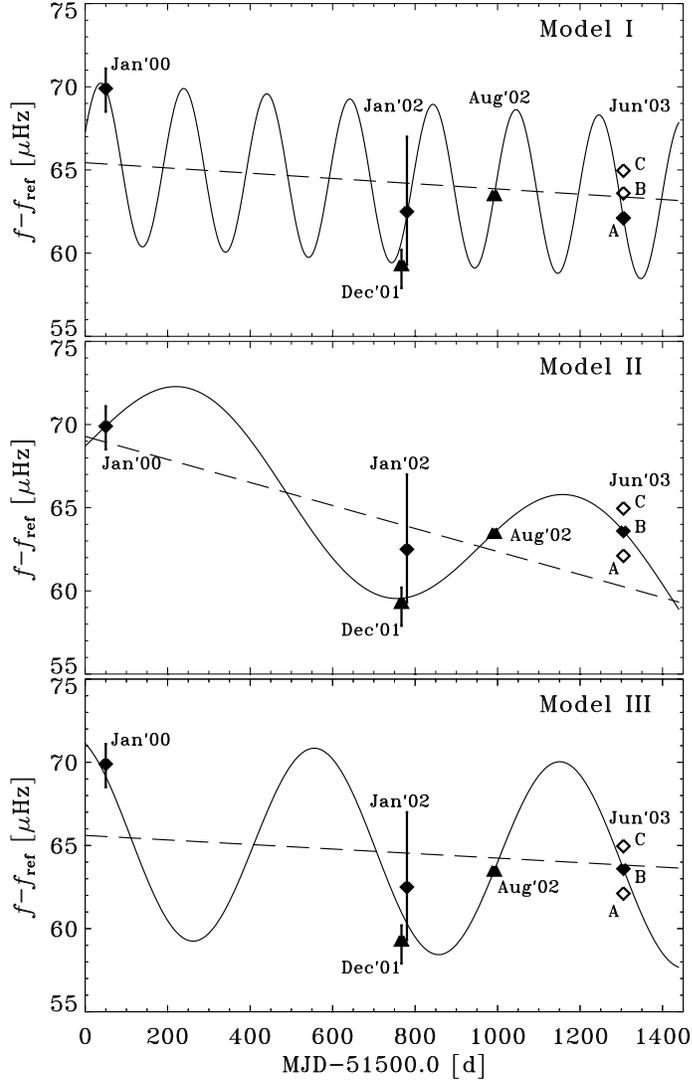}
\caption{
The same timing points as in Figure 3 are fitted with
three models of binary motion (see \S3.3 for the model parameters).
The dashed lines show
the steady spin-down components in the binary-model fits.
The filled diamonds for the June 2003 observation indicate the frequency
values chosen for each of the models
at that epoch (see Fig.\ 1 and the text for details).
}
\end{figure}
\end{document}